\documentclass[showpacs,twocolumn,amsmath,amssymb,pre,aps]{revtex4}
\usepackage{graphicx}
\usepackage{bm}
\usepackage{mathrsfs}

\begin{document}
\title
{Dispersion relations for circular single and double dusty plasma chains}
\draft

\author{D.~V.~Tkachenko$^{1}$, T.~E.~Sheridan$^{2}$ and V.~R.~Misko$^{1}$}

\affiliation
{$^{1}$Department of Physics, University of Antwerpen, Groenenborgerlaan 171, B-2020 Antwerpen, Belgium\\
$^{2}$Department of Physics and Astronomy, Ohio Northern University, Ada, Ohio 45810, USA}

\date{\today}

\begin{abstract}
We derive dispersion relations for a system of identical particles confined in a two-dimensional annular harmonic well and which interact through a Yukawa potential, e.g., a dusty plasma ring.  When the particles are in a single chain (i.e., a one-dimensional ring) we find a longitudinal acoustic mode and a transverse optical mode which show approximate agreement with the dispersion relation for a straight configuration for large radii of the ring. When the radius decreases, the dispersion relations modify: there appears an anticrossing of the modes near the crossing point resulting in a frequency gap between the lower and upper branches of the modified dispersion relations. For the double chain (i.e., a two-dimensional zigzag configuration) the dispersion relation has four branches: longitudinal acoustic and optical and transverse acoustic and optical.
\end{abstract}


\maketitle

\section{Introduction}
An interest in the structural and dynamical properties of ordered many-body systems confined by various geometries is due to the recent advances in studies of, e.g., confined plasma \cite{PS_80, PRE_81, PRL_104, Lowen}.
Knowledge of the properties, e.g., the ground state configurations and the normal mode spectra of the system allows one to calculate thermodynamic characteristics such as the free energy, entropy, etc.
For example, theoretical investigation of properties of one- and quasi-one dimensional linear infinite chains of particles were presented in Refs. \cite{PRE_70, PRB_69, PRB_77, PRB_81}.
Experimental studies of vibrational properties for one-dimensional linear finite chains of particles were performed in Refs. \cite{PRL_91, PRE_71, PRE_81}.

The ground state configurations and the normal mode spectra of a quasi-one-dimensional multichain plasmas of charged particles interacting through a screened Coulomb potential were investigated for monodispersed particles \cite{PRB_69, PRE_70} and for binary mixtures \cite{PRB_77}.
The one- to two-dimensional continuous zigzag transition was studied experimentally \cite{PRE_81} for small clusters of particles interacting through a Yukawa potential, and theoretically in \cite{PRB_81} for confined along one direction systems of strongly interacting particles.
Continuous and discontinuous structural phase transitions in confined systems were also investigated \cite{PRB_69, PRB_77}.
The evidence of a direction-dependent melting and reentrant melting as a function of the density of particles was found in \cite{PRB_69, PRB_49}.
In the experiment \cite{PRL_91} a transverse optical mode in a one-dimensional Yukawa chain which had negative dispersion (i.e., when phase and group velocities were oppositely directed) was observed.
The dynamic structure factor of one-dimensional chains with different types of disorder was studied in \cite{Helbig_Just, Engel_Sonntag}.
The linear and weak non-linear wave propagation in one-dimensional lattices of dust plasma was investigated in Ref. \cite{Melandso}.
The studies of the energy spectrum, the eigenmodes, the density of states and melting properties of finite two-dimensional Yukawa lattices confined in parabolic and/or hard-wall external potential were performed in Refs. \cite{PRB_51, PRB_49, PRE_64, PRE_72}.

The properties of two-dimensional Coulomb clusters confined in a harmonic potential were theoretically studied in Ref. \cite{PRB_54}.
A numerical analysis of normal modes of the ground-state configuration was performed, and the stability of the ring cluster was analyzed.
In \cite{PS_80} one-dimensional and quasi-one-dimensional strongly-coupled dusty plasma rings were created experimentally, and the longitudinal and transverse dispersion relations for the one-dimensional single chain were measured.
The measured dispersion relations were in a good agreement with theoretically predicted results for one-dimensional ring

In this work, we theoretically investigate the normal mode spectra for single and double circular chains.
We analyze these spectra as a function of varying parameters such as the radius of the channel $\sigma$ and the density of particles $n=N/2\pi r_0$, where $r_0$ is the equilibrium radius of the chain and $N$ is the number of particles in the chain.
In addition, we compare the obtained spectra for a double circular chain with those for a single chain (i.e., in the vicinity of the zig-zag transition) and for a double straight chain (i.e., for large radii of the circular chain).

The paper is organized as following.
First we discuss the normal mode spectra of a single circular chain in Sec. II.
Then in Sec. III we derive the dispersion relations for an infinite double straight chain and finally, we present the normal mode theory for the case of double circular chain in Sec. IV.
Conclusions are given in Sec. V.

\section{Single circular chain}

Let us first consider a system of $N$ particles in a circular parabolic confinement which interact via a Yukawa potential. Fisrt, we discuss the case of a single chain.
The potential energy of the system in dimensionless form is:
\begin{equation}\label{eqUC}
U=\sum_i \left(\sigma - r_i\right)^2 + \sum_{i<j} \frac{e^{-\kappa r_{ij}}}{r_{ij}},
\end{equation}
where $\sigma$ is the radius of the circular channel, $r_i$ is the coordinate of $i$th particle, $r_{ij}$ is the interparticle distance, and $\kappa$ is the inverse Debye screening length. The transformation to dimensionless units is defined by the following relations: $r_{ij}=\tilde{r}_{ij}/r_o,\ \sigma_{ij}=\tilde{\sigma}/r_o,\ r_i=\tilde{r}_i/r_o,\ \kappa=\tilde{\kappa}r_o,\ U=\tilde{U}/U_o$, and finally $U_o^3=\alpha q^4/\epsilon^3,\ r_o^3=q^2/\alpha\epsilon$, where all the dimensional units are labeled with tilde, and $\alpha$ is the strength of the circular parabolic confinement, $q$ is the charge of a particle.
The equilibrium chain configurations required for the calculation of the normal modes are found by minimizing the potential energy of the system.

\begin{figure}[btp]
\begin{center}
    \includegraphics*[width=7.0cm]{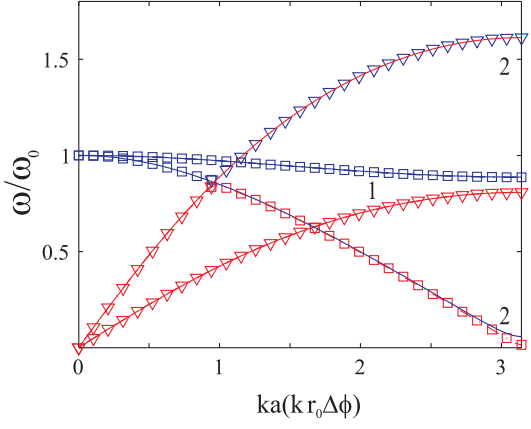}
    \caption{
	The dispersion curves for single circular (triangles for longitudinal modes and squares for transverse modes) and straight (lines) chains for two different values of density $n=0.596$ (curves 1) and $0.892$ (curves 2) (cp. Ref. \cite{PS_80}) and for the circle radius $\sigma=16$. Dashed lines are guides for eye. The same symbols are used in Figs.~2 and 3.
	}
\end{center}
\label{FigI}
\end{figure}

The Lagrangian of this system in polar coordinates is \cite{Landau}:
\begin{eqnarray}
\label{eqLS}
L 
& = & 
\frac{m}{2}\sum_{n}\left(\left(\dot{q}_{n}^{r}\right)^2 + \left(r_{0} \dot{q}_{n}^{\phi}\right)^2\right) 
\\ \nonumber 
& - & 
\frac12 \sum_{\alpha,\beta,n-n'} \lambda^{\alpha\beta}\left(n-n'\right) q_{n}^{\alpha} q_{n'}^{\beta},
\end{eqnarray}
where $(\alpha,\beta)=(r,\phi)$ are the coordinate indices, $(n, n')=1..N/2$ are the indices of the cells, $q_{n}^r=r_{n}-r_{0},\ q_{n'}^{\phi}=\phi_{n'}-\phi_{n'0}$ are the coordinates, and $r_0$ is the equilibrium radius of the chain.
The coefficients of the elastic coupling are given by:
\begin{equation}\label{eqMeS}
\lambda^{\alpha\beta}\left(n-n'\right)=\partial_{q_n^\alpha}\partial_{q_{n'}^\beta}U|_{eq},
\end{equation}
where the $eq$ denotes the set of the equilibrium values of all coordinates ${r_{n0},\phi_{n0}}$.
Solving the problem of normal modes, we obtain:
\begin{equation}\label{eqIIS}
\Lambda=
\left(\begin{array}{cccc}
\Lambda^{\phi\phi}&\Lambda^{\phi r}\\
\Lambda^{r\phi}&\Lambda^{rr}\\
\end{array}\right)
=m\omega^2
\left(\begin{array}{cccc}
r_0^2&	0\\
0& 	1
\end{array}\right),
\end{equation}
where the matrix elements are defined by
\begin{equation}\label{eqMS}
\Lambda^{\alpha\beta}=\frac12\sum_{j}\lambda^{\alpha\beta}\left(j\right)e^{ik r_0 j\Delta\phi}.
\end{equation}
Here $j=n-n'$, $\Delta\phi=2\pi/N$, $k=0..\pi/a$ is the wave vector, where $a=r_0\Delta\phi$ is the distance between nearest neighbours.

The solution of the problem of normal modes in polar coordinates results in the following general dispersion relation for a circular single chain of particles:
\begin{widetext}
\begin{eqnarray}
\label{eqSC}
\frac{\omega^2}{\omega_0^2}
& = & 
\frac12+\frac12\sum_{j=1}^{\left[N/2\right]}
\frac{e^{-\kappa r}}{r^3}\left(\kappa^2 r^2 + \kappa r + 1\right)|_{r=r_j}\cos\left(j\Delta\phi\right)\left(1-\cos\left(k r_0 j\Delta\phi\right)\right)
\\ \nonumber
& \pm & 
\frac12\left[\left(1-\sum_{j=1}^{\left[N/2\right]}
\frac{e^{-\kappa r}}{r^3}\left(\kappa^2 r^2 + 3\kappa r + 3\right)|_{r=r_j}\left(1-\cos\left(k r_0 j\Delta\phi\right)\right)\right)^2  \right. 
\\ \nonumber
& + & 
\left. 
\left( \sum_{j=1}^{\left[N/2\right]}\frac{e^{-\kappa r}}{r^3}\left(\kappa^2 r^2 + \kappa r + 1\right)|_{r=r_j}\sin\left(j\Delta\phi\right)\sin\left(k r_0 j\Delta\phi\right)\right)^2\right]^{\frac12} \nonumber
\end{eqnarray}
\end{widetext}
where $r_j=2r_0\left|\sin\left(\frac{j\Delta\phi}{2}\right)\right|$ and $\Delta\phi$ is the angular distance between nearest neighbours.

In the limiting case of large radius $r_0\rightarrow \infty\ (\Delta\phi\rightarrow 0)$
Eq.~(\ref{eqSC}) can be reduced to the analytical result for the single straight chain (cp. Ref. \cite{PRE_70}):
\begin{eqnarray}\label{eqSS}
\frac{\omega_T^2}{\omega_0^2}=1-\sum_{j=1}^{\infty}
\frac{e^{-\kappa r}}{r^3}\left(\kappa r+1\right)|_{r=r_j} \left(1-\cos\left(kja\right)\right),\\
\frac{\omega_L^2}{\omega_0^2}=\sum_{j=1}^{\infty}
\frac{e^{-\kappa r}}{r^3}\left(\left(\kappa r+1\right)^2 + 1\right)|_{r=r_j} \left(1-\cos\left(kja\right)\right).
\end{eqnarray}
where $r_j=ja$.

Relation (\ref{eqSC}) can be further simplified if we consider nearest neighbours approximation (NNA).
In this case, one should omit all the terms in the sums of Eq.~(\ref{eqSC}) except for
the term $j=1$ which defines the elastic coefficient $\chi$ between the {\it nearest} particles.
Then we obtain:
\begin{widetext}
\begin{equation}\label{eqSCNNA}
\frac{\omega^2}{\omega_0^2}=1 + \chi\cos\left(\Delta\phi\right)\left(1-\cos\left(k\Delta\phi\right)\right)\pm
\sqrt{\left(1 - \chi\left(1 -\cos\left(k\Delta\phi\right)\right)\right)^2 + \chi^2\sin^2\left(\Delta\phi\right)\sin^2\left(k\Delta\phi\right)}.
\end{equation}
\end{widetext}

\begin{figure}[btp]
\begin{center}
    \includegraphics*[width=7.0cm]{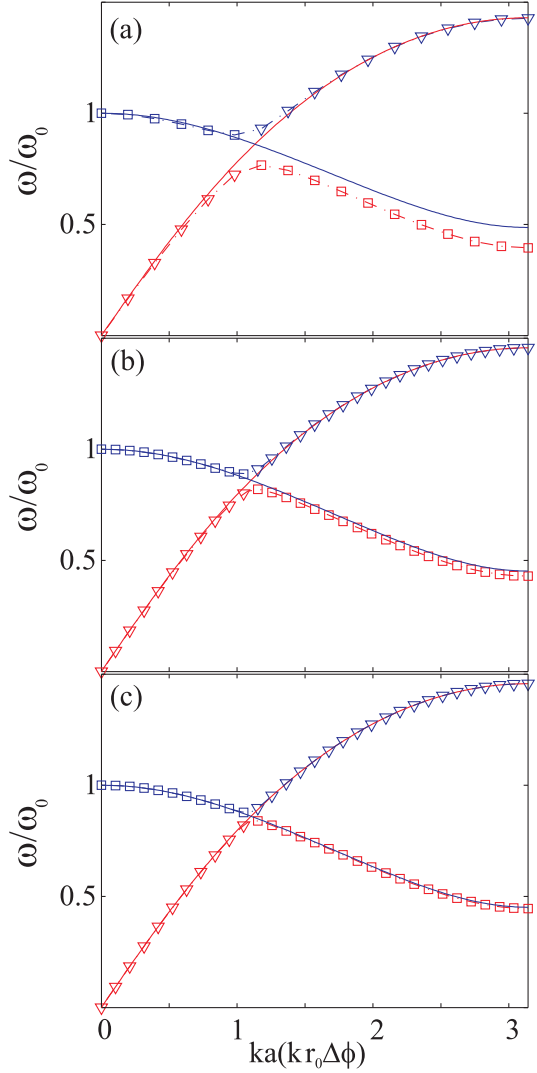}
    \caption{The dispersion curves for single circular (open circles) and straight (lines) chains for density $n=0.84$, for various values of the radius $\sigma=2.94$ (a), $6.0$ (b) and $12.09$ (c).}
\end{center}
\label{FigIII}
\end{figure}

\begin{figure}[btp]
\begin{center}
  \includegraphics*[width=7.0cm]{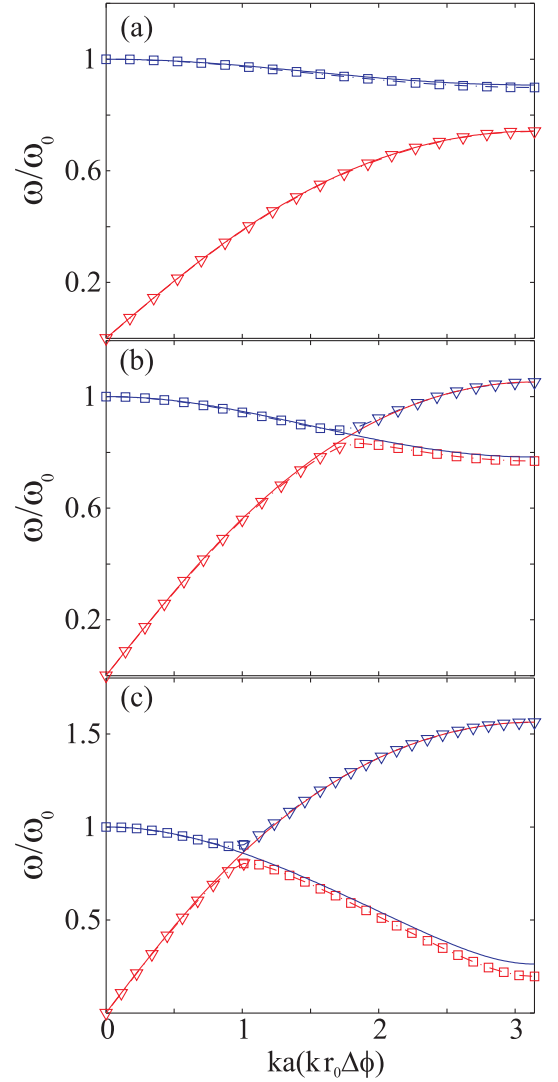}
  \caption{The dispersion relations for circular (open circles) and straight (lines) chains for a channel of radius $\sigma=5$ and for different values of the density $n=0.57$ (a), $0.69$ (b) and $0.88$ (c).}
\end{center}
\label{figIV}
\end{figure}

The dispersion curves calculated using Eq.~(\ref{eqSC}) are presented in Fig.~1 for the same parameters (for densities $n=0.596;0.892$ and corresponding numbers of particles $N=60;90$, the radius of a parabolic channel is $\sigma=16$) for single circular (open circles) and straight (lines) chains.
Results were checked by computing normal mode spectra using the method of Ref.~\cite{PS_80} and found to be in good agreement with the calculations presented here.
It is worth noting that the agreement between the dispersion curves is fairly good not only for large chains but remarkably even for chains with rather small radii as it can be seen from Fig.~2.
The small difference in the dispersion curves for a circular chain and for a straight chain is related to the fact that there is a non-zero coupling between the directions of oscillations in the transverse (radial) and the longitudinal (tangential) modes in case of a circular chain (Fig.~1 and Fig.~2).
For a system with non-zero curvature these modes are not independent unlike for the case of a straight chain.
In order to clarify the dependence of the dispersion curves on the curvature of the channel, we show in Fig.~2 these curves for different radii $\sigma$ but for the same density $n=0.84$.
As it can be seen from Fig.~2, the difference between the curves for circular and straight chains diminishes when the radius of the chain increases.
The main qualitative difference is observed for intermediate $k$-values while for small and large $k$ the dispersion curves are approximately identical.
Although we should also note that there is a shift of the longitudinal branch in the area of intermediate and large $k$ which increases with decreasing the channel radius $\sigma$.
In particular, we observe the appearance of a ``gap'' between the transverse and longitudinal branches of the dispersion curves (``anticrossing'') for a circular chain.
Note that the modes are discrete (see, e.g., Fig.~2(a)) and therefore this gap is well defined only for large enough density $n$ when the dispersion curves become quasi-continuous.
It is also seen from Fig.~2 that the width of the gap increases with decreasing radius of the circular chain $r_0$. Inside the frequency gap no oscillations can be excited in the system.

Now we will discuss the dispersion curves as a function of the density of particles $n$ for small channel radius $\sigma=4$.
The corresponding plots are shown in Fig.~3.
Figs.~3(a) refers to low densities $n$ which is equivalent to the situation of weak interparticle interaction.
The distance between the nearest particles is large enough, therefore in spite of the strong coupling between transverse and longitudinal modes of the dispersion curves (shown by symbols in Fig.~3(a)) are similar to the curves for a straight single chain (shown by lines).
For higher densities $n$ (Figs.~3(b)-(c)) these differences become more pronounced.
For high density $n$, particles are situated closer to each other and therefore they interact stronger.
This results in an anticrossing of the transversal and longitudinal dispersion branches (lines).
Due to the circular symmetry of the system, there is a coupling between transversal and longitudinal modes that leads in turn to the splitting of the transverse and longitudinal branches.
This coupling is absent in case of a straight chain therefore there is no gap in the normal mode frequencies (see Fig.~3 (lines)).
It should also be noted that the transverse branch gradually approaches zero at the edge of the band.
As soon as the longitudinal frequency approaches zero, a phase transition to the double-chain configuration takes place \cite{PS_80}.
The condition for the zig-zag phase transition was discussed in \cite{PS_80, PRB_81}.

\section{Straight double chain}

The dispersion properties of a quasi-one-dimensional classical Wigner crystal were investigated in detail in Ref.~\cite{PRE_70} in the presence of dissipation.
The results without dissipation were analyzed in Ref.~\cite{PRB_69} and corresponding result for the binary system of particles in \cite{PRB_77}.
Nevertheless it is worth to revisit here the dispersion relations for a multichain case.

Let us consider a system of particles confined to a straight channel.
The potential energy of the system in dimensionless form is described by:
\begin{equation}\label{eqUI}
U=\sum_i y_i^2 + \sum_{i<j} \frac{e^{-\kappa r_{ij}}}{r_{ij}}.
\end{equation}
Solving the problem of normal modes for this system results in the following dynamical matrix:
\begin{equation}\label{eqI}
\Lambda=
\left(\begin{array}{cccc}
\Lambda_{11}^{xx}&	0&			\Lambda_{12}^{xx}&	\Lambda_{12}^{xy}\\
0& 			\Lambda_{11}^{yy}&	\Lambda_{12}^{yx}&	\Lambda_{12}^{yy}\\
\Lambda_{21}^{xx}&	\Lambda_{21}^{xy}&	\Lambda_{22}^{xx}&	0\\
\Lambda_{21}^{yx}&	\Lambda_{21}^{yy}& 	0&			\Lambda_{22}^{yy}
\end{array}\right)
=m\omega^2
\left(\begin{array}{cccc}
1& 0& 0& 0\\
0& 1& 0& 0\\
0& 0& 1& 0\\
0& 0& 0& 1
\end{array}\right)
\end{equation}
where the matrix elements are given by
\begin{equation}\label{eqL}
\Lambda_{ls}^{\alpha\beta}=\frac12\sum_{j}\lambda_{ls}^{\alpha\beta}\left(j\right)e^{ikja},
\end{equation}
where $j$ denotes the $j$th elementary cell of the chain, $(l,s)=(1,2)$ is the index of the chain, $a$ is the interparticle distance (in the same chain) and $(\alpha,\beta)=(x,y)$ are the coordinates.

The independent matrix elements of the dynamical matrix $\Lambda$ are given in Appendix A.
Other elements of the dynamical matrix (\ref{eqI}) can be defined using the following relation from the theory of normal modes \cite{Landau}:
\begin{equation}\label{eqHerm}
\Lambda=\Lambda^\dagger.
\end{equation}
It is clear that the presented results reveal some differences with similar results obtained in \cite{PRE_70, PRB_69} for a straight zigzag configuration.
Let us briefly discuss them.
Coefficient $\lambda_{ls}^{\alpha\beta}$ defines the elastic coupling between the $\alpha$ coordinate of some particle of the $l$th chain from zeroth cell and $\beta$-coordinate of a particle of s-$th$ chain from $j$th cell (see Fig.~4).
(Further we will call these coefficients ``$\alpha\beta$-couplings''.)
As it follows from Refs.~\cite{PRE_70, PRB_69} all $\alpha\beta=xy(yx)$ couplings are equal to zero.
Our calculations give the same result only for particles of the {\it same} chain, i.e., when $l=s$.
However, in the general case $l\neq s$ these matrix elements are non-zero.
This can be demonstrated by a direct calculation and be explained from physical arguments.

\begin{figure}[btp]
\begin{center}
    \includegraphics*[width=8.0cm]{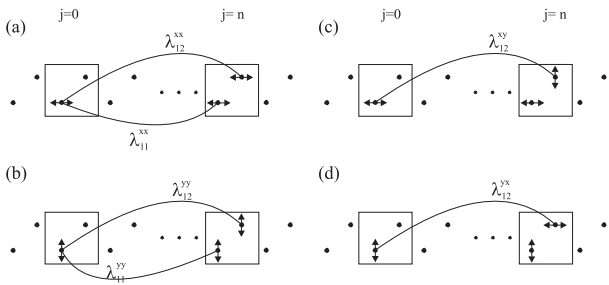}
    \caption{Scheme illustrating elastic couplings in a straight double chain. Coefficients $\lambda_{ls}^{\xi\zeta}$ describe coupling between $\xi$-oscillations of $l$th chain and $\zeta$-oscillations of $s$th chain. Plots (a) and (b) show the couplings between longitudinal (a) and between transverse (b) of the same or different chains. Plots (c) and (d) show the couplings between the longitudinal and transverse oscillations in different chains [see Eq.~(\ref{eqI})].}
\end{center}
\label{FigII}
\end{figure}

First, we discuss the case of $\alpha\beta$-couplings for $l=s$ (see Figs.~4(a)-(b) for $\lambda_{11}^{xx}$ and $\lambda_{11}^{yy}$).
If some oscillations are excited in the $y$-direction in the system then this would
obviously result in a change of the length of couplings (and as a result, in its strength) with other particles and,
as a consequence, this induces oscillations of other particles in the $x$-direction (see Fig.~4(a)-(b)).
However, within the harmonic approximation this variation of coupling strength is of second order in the strength of coupling or first order in the $y$-deviation.
Thus we can conclude that the oscillations cannot be excited in this case as verified by direct calculations.
The application of the same procedure to the $yx$-coupling leads to a similar result.

However the situation is different for the case $l\neq s$ (see Figs.~4(c)-(d) for $\lambda_{12}^{xy}$ and $\lambda_{12}^{yx}$).
In this case, variations of the strength of the $\alpha\beta$-coupling between particles from different chains (which take place for, e.g., transverse oscillations) are of the same order as the deviation from equilibrium.
A displacement along the $y$-direction of a particle results in the corresponding displacement of the nearest particles of the adjacent chain in the
$x$-direction (and vice versa).
It is reflected by the fact that the corresponding coefficients ($\lambda_{12}^{xy}$ and $\lambda_{12}^{yx}$) are non-zero, although the absolute values of these coefficients are smaller than those for the $xx-$ and $yy$-couplings.

\begin{figure}[btp]
\begin{center}
 \includegraphics*[width=8.0cm]{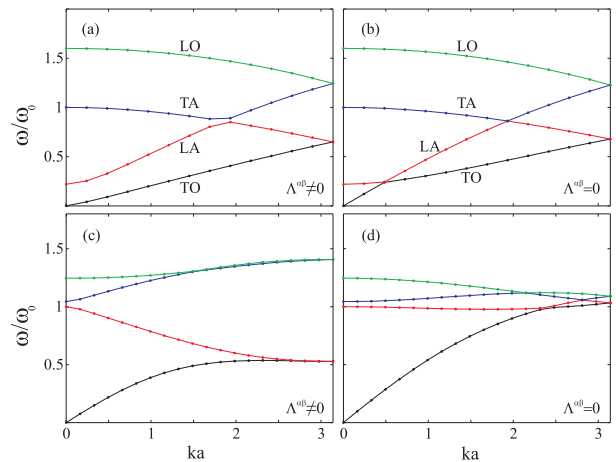}
  \caption{The dispersion curves for a double straight chain for various values of density $n$ and the equilibrium interchain distance $d$: $n=0.90; d=0.209; \sigma=4.5; N=26$ (a)-(b) and $n=1.19; d=0.748; \sigma=5.0; N=38$ (c)-(d). The dependencies were calculated using relations (\ref{eqI}), (\ref{eqI2})-(\ref{eqI6}) [(a) and (c)] cases and using the same relations but with zero second diagonal in the dynamical matrix \cite{PRB_69} [(b) and (d)].}
\end{center}
\label{FigV}
\end{figure}

Therefore the results presented in Refs.~\cite{PRE_70, PRB_69} are valid only for a limiting case of a very narrow double chain (i.e., when $d<<a$).
In this case the $\alpha\beta$-couplings for $l=s$ become approximately applicable to the case $l\neq s$.
This remark remains valid for all other multichain configurations considered in Refs.~\cite{PRE_70, PRB_69}.
Thus presented formulas (\ref{eqI}), (\ref{eqI2})-(\ref{eqI6}) converge to the ones obtained in Refs.~\cite{PRB_69, PRE_70} when the condition $\frac{d}{a}\ll1$ is fulfilled.
In this case formulas (\ref{eqI}), (\ref{eqI2})-(\ref{eqI5}) are valid and $\lim_{d\ll a}\Lambda_{ls}^{\alpha\beta}|_{\alpha\neq\beta,l\neq s}=0$.

Let us now discuss the dispersion relations for a double straight chain presented in Fig.~5.
As shown in Figs.~5(a)-(b) for small ratio $\frac{d}{a}=0.081$ ($\frac{d}{a}\ll 1$) the curves calculated using Eqs. (\ref{eqI}), (\ref{eqI2})-(\ref{eqI6}) and formulas from Ref.~\cite{PRB_69} look similar.
Fig.~5(b) shows that the LA-branch (longitudinal acoustic) of vibrations intersects both the transverse branches (TA and TO).
This fact is a consequence of neglecting the interaction between the transverse and longitudinal vibrations in the dynamic matrix.
Taking into account the interaction between transverse and longitudinal vibrations leads to the ``anticrossing'' in the crossing points of LA and TO, LA and TA branches as shown in Fig.~5(a).
As a result, the LA, TA and TO branches change and acquire frequency gaps.
On the other hand, the difference between these two cases (i.e., with and without the $xy-$coupling taken into account) becomes more pronounced when the ratio $\frac{d}{a}$ is not small: $\frac{d}{a}=0.389$ (Figs.~5(c)-(d)).
From Fig.~5(c) it is seen that only the LO and TA branches experience hardening (unlike in Fig.~5(d) where all the branches except the LA branch show only a weak dependence on the wave vector).
Fig.~5(c) shows that for density $n=1.19$ there is a narrow frequency gap between the longitudinal and transverse branches.

Thus the account of the interaction between the transverse and longitudinal oscillations for different chains in case of a circular double chain is important and leads to an ``anticrossing'' in the corresponding dispersion branches (Figs.~5(a),(c)).

\section{Circular double chain}
\subsection{Dispersion relations}

\begin{figure}[btp]
\begin{center}
 \includegraphics*[width=8.0cm]{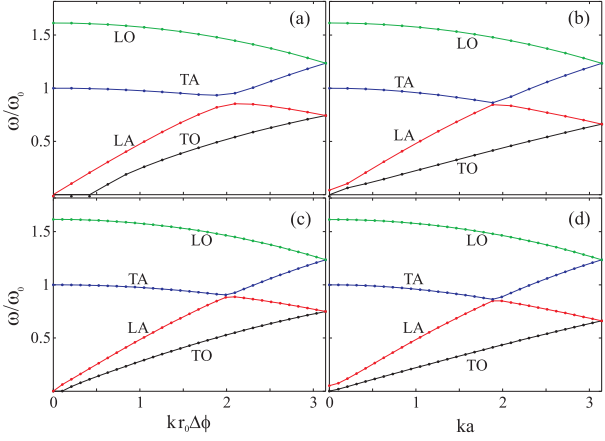}
  \caption{The dispersion curves for double circular (the left column) and straight (the right column) chains for various values of channel radius but for the same density $n=0.934$ close to phase transition point. The channel radius and the number of particles is $\sigma=5.26$, $N=30$ - (a) and $\sigma=26.66$, $N=150$ - (c).}
\end{center}
\label{FigVI}
\end{figure}

\begin{figure}[btp]
\begin{center}
 \includegraphics*[width=8.0cm]{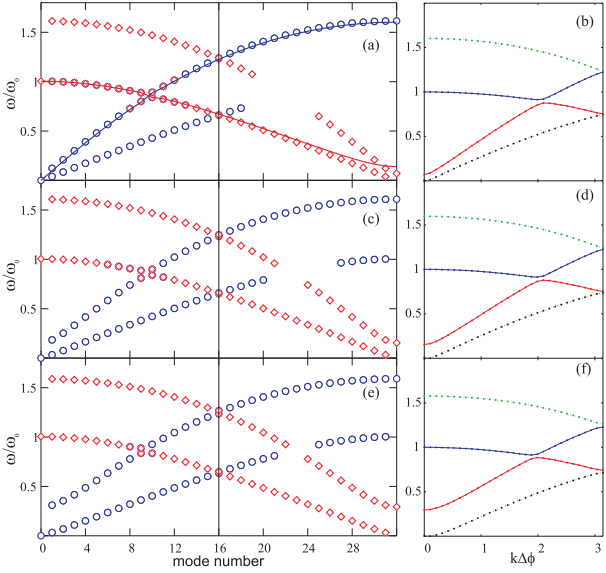}
  \caption{The dispersion curves for double circular chain for various values of the channel radius $\sigma=11.33$ (a), (b); $11.30$ (c), (d) and $11.20$ (e), (f) near the single-to-double ring transition point. Comparison of the dispersion curves obtained analytically using Eq.~(\ref{eqII}) (right column) and those calculated numerically (left column). The number of particles for all cases is the same: $N=64$, and the corresponding densities are $n=0.896$ (a), (b); $0.898$ (c), (d) and $0.906$ (e), (f).}
\end{center}
\label{FigVII}
\end{figure}

Next we study the dispersion properties of a double circular chain.
The potential energy of the system is given by (\ref{eqUC}).
Due to the symmetry of this system the Lagrangian can be written in polar coordinates in a following manner \cite{Landau}:
\begin{eqnarray}
\label{eqL}
L 
& = & 
\frac{m}{2}\sum_{l,n}\left(\left(\dot{q}_{ln}^{r}\right)^2 + \left(r_{l0} \dot{q}_{ln}^{\phi}\right)^2\right) 
\\ \nonumber 
& - & 
\frac12 \sum_{s,l,\alpha,\beta,n,n'} \lambda_{sl}^{\alpha\beta}\left(n-n'\right) q_{sn}^{\alpha} q_{ln'}^{\beta},
\end{eqnarray}
where $(l,s)=(1,2)$ are the indices of the chains, $(\alpha,\beta)=(r,\phi)$ are the coordinate indices, $q_{sn}^r=r_{sn}-r_{s0},\ q_{ln'}^{\phi}=\phi_{ln'}-\phi_{ln'0}$ are the coordinates, $r_{l0}$ is the equilibrium radius of the $l$th chain, and $(n,n')=1..N/2$ are the indices of the cells.
The coefficients of the elastic coupling are given by:
\begin{equation}\label{eqMe}
\lambda_{sl}^{\alpha\beta}\left(n-n'\right)=\partial_{q_{sn}^\alpha}\partial_{q_{ln'}^\beta}U|_{eq},
\end{equation}
where $eq$ denotes the set of coordinates ${r_{sn0},\phi_{sn0}}$ at their equilibrium value.
Solving the problem of normal modes, we obtain:
\begin{equation}\label{eqII}
\Lambda=
\left(\begin{array}{cccc}
\Lambda_{11}^{\phi\phi}&\Lambda_{11}^{\phi r}&\Lambda_{12}^{\phi\phi}&\Lambda_{12}^{\phi r}\\
\Lambda_{11}^{r\phi}&	\Lambda_{11}^{rr}&	 \Lambda_{12}^{r\phi}& \Lambda_{12}^{rr}\\
\Lambda_{21}^{\phi\phi}&\Lambda_{21}^{\phi r}&\Lambda_{22}^{\phi\phi}&\Lambda_{22}^{\phi r}\\
\Lambda_{21}^{r\phi}&	\Lambda_{21}^{rr}&	 \Lambda_{22}^{r\phi}& \Lambda_{22}^{rr}
\end{array}\right)
=m\omega^2
\left(\begin{array}{cccc}
r_{10}^2&	0&	0&		0\\
0& 		1&	0&		0\\
0&		0&	r_{20}^2&	0\\
0&		0&	0&		1
\end{array}\right)
\end{equation}
where the matrix elements are defined by
\begin{equation}\label{eqM}
\Lambda_{ls}^{\alpha\beta}=\frac12\sum_{j}\lambda_{ls}^{\alpha\beta}\left(j\right)e^{ik r_0 j\Delta\phi}.
\end{equation}

Here $j=n-n'$, and $l,\ s$ are defined above, $N$ is the number of particles, $\Delta\phi=4\pi/N$ and $k$ is the wave vector.
The independent matrix elements of the dynamical matrix $\Lambda$ are given below in Appendix B.
Taking into attention the asymptotic relations given in Appendix B in the limit $r_{l0}\rightarrow\infty,\ (\Delta\phi\rightarrow0)$ and $\frac{d}{\sigma}\rightarrow0$ we obtain, as a consequence, the following relations:
\begin{widetext}
\begin{eqnarray}\label{eqII9}
\lim\Lambda_{ll}^{rr}=\Lambda_{ll}^{yy},\ \ \lim\Lambda_{ll}^{r\phi}=\Lambda_{ll}^{yx}=0,\ \
\lim\Lambda_{ls}^{rr}=\Lambda_{ls}^{yy}=0,\ l\neq s \ \
\lim\Lambda_{ls}^{r\phi}=\Lambda_{ls}^{yx}=0,\ l\neq s \\ \nonumber
\lim\frac{1}{r_{l0}^2}\Lambda_{ll}^{\phi\phi}=\Lambda_{ll}^{xx},\ \
\lim\frac{1}{r_{l0}^2}\Lambda_{ls}^{\phi\phi}=\Lambda_{ls}^{xx}=0,\ l\neq s \ \
\lim\frac{1}{r_{l0}^2}\Lambda_{ls}^{\phi r}=\Lambda_{ls}^{xy}=0,\ l\neq s.
\end{eqnarray}
\end{widetext}

By direct calculation one can show that the obtained formulas (\ref{eqII}), (\ref{eqII1})-(\ref{eqII7}) converge to the ones found in the previous section, (\ref{eqI}), (\ref{eqI2})-(\ref{eqI6}) [in the limit defined by expressions (\ref{eqII8}) and (\ref{eqII9})], and in the limiting case $d<<a$ to the corresponding results of \cite{PRB_69}.
This  is also verified by a comparison of the dispersion curves for double circular (left column) and straight (right column) chains which are presented in Fig.~6.
The dispersion curves were calculated for various values of channel radius $\sigma$ but for the same density $n\approx0.934$ close to phase transition point.
The channel radius and the number of particles are $\sigma=5.26, N=30$ for (a) and $\sigma=26.66, N=150$ for (c) plots.
Comparison of Fig.~6(a) and (b) shows the presence of a gap in the dispersion relations for a circular chain and a straight chain.
In addition, as we can see from Fig.~6(a), the LA-branch for small $k$ ($k\approx0$) is characterized by $\omega\approx0$ which means
that in the vicinity to $k = 0$ the longitudinal mode is pure rotational and it requires zero energy to excite.
Comparison of Fig.~6(c) and (d) reveals significant similarity between the dispersion curves for double circular and straight chains.
This is reasonable because this case corresponds to small curvature.
Thus we can conclude that differences between the dispersion relations of the circular and straight chains increase with decreasing channel radius $\sigma$.

\subsection{The single- to double-chain transition}

Here we analyze the special case of the single-to-double chain transition for a circular chain.
For that purpose, we considered the limit $d\rightarrow0$ in the expressions for a double circular chain.
On the other hand, in order to verify the obtained results for this limit, we performed an independent numerical calculation for a
double circular chain just above the single-to-double transition.
The results obtained using both methods are shown in Fig.7 for $\sigma\approx11.3$ and $n\approx0.9$.
Comparison of numerical and analytical calculations reveals an excellent agreement (see Fig.~7).

Note that at the zig-zag transition point,
the length of the unit cell increases from $a$ to approximately $2a$ (for each chain in the double-ring configuration). Therefore, the Brillouin zone shrinks (left column) giving rise to the new brunches in the dispersion relations.

\subsection{Role of the density}

\begin{figure}[btp]
\begin{center}
 \includegraphics*[width=8.0cm]{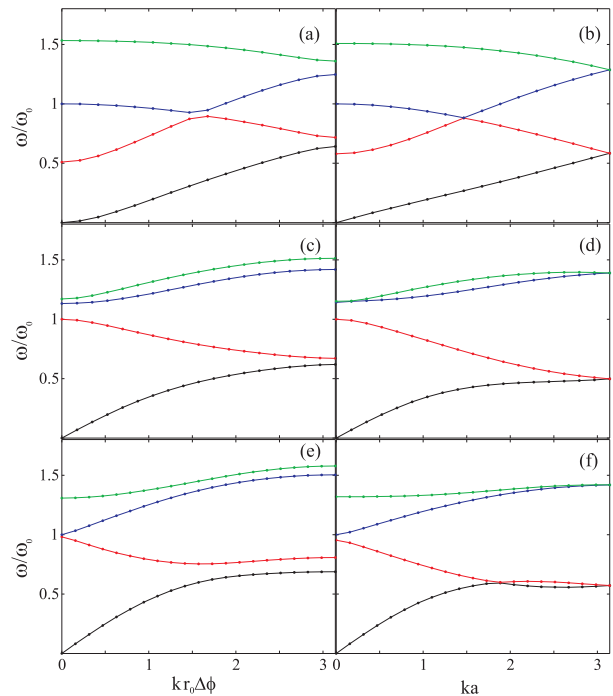}
  \caption{The dispersion curves for double circular (left column) and straight (right column) chains for various values of density $n$ beyond phase transition point (0.94(N=30)-(a),(b); 1.13(N=36)-(c),(d); 1.25(N=40)-(e),(f)) for channel radius $\sigma=5.0$.}
\end{center}
\label{FigVIII}
\end{figure}

The dispersion relations are shown in Fig.~8 for the same channel radius $\sigma=5.0$ for double circular channel (left column) calculated using Eqs.~(\ref{eqII}), (\ref{eqII1})-(\ref{eqII7}) and straight (right column) calculated using Eqs.~(\ref{eqI}), (\ref{eqI2})-(\ref{eqI6}) for densities $n$ slightly above the critical density $n_c$ at the single-to-double chain transition.
As one can see from Fig.~8, difference between the dispersion curves for double circular and straight chains becomes more pronounced with increasing particle density $n$.
Therefore, the best agreement between the dispersion curves for double circular and straight chains is reached for the lowest value of particle density $n$ just above the critical density at the phase transition point.
However this fact is not obvious: it is clear that the best correspondence between the dispersion curves for double circular and double straight chains is reached when the unit cells for both chains are very similar.
On the other hand, increase of the density $n$ results in enhancing this similarity (due to decreasing angular spacing $\Delta\phi$) and  simultaneously in its diminishing (due to increasing interchain spacing $d$).
Therefore we studied the imbalance between the ``lattice constants'' in the inner and outer chains, $a_2-a_1=d\Delta\phi$ as a function of the density $n$ and we found that the maximum similarity in the unit cells for double circular and straight chains is indeed reached for critical density $n_c$.

From Fig.~8(a), (c) and (e) it can be seen that the gap between dispersion curves first increases with growing density and then decreases.
In addition the frequency of TO branch experiences hardening for small $k\approx0$ as it is seen from comparison of Fig.~8(a) and (c).

Thus Fig.~8 reveals that there is essential coupling between LA and TO and LO and TA branches of the eigenmodes.

\section{Conclusions}

We derived an analytical expression for the dispersion relations for a single circular chain of interacting particles and analyzed the obtained results for different radii of the chain and the influence of the curvature of the confinement channel.
For large radius of the chain our results converge to the dispersion relations of a straight single chain \cite{PRE_70}.

An improved expression for the dynamical matrix for the case of a straight double chain was obtained that takes into account all the elastic (harmonic) couplings between the oscillations of the particles in the different chains.
Taking into account the interaction between the transverse and longitudinal oscillations of the different chains for the case of a double straight chain leads to a weak transverse-longitudinal ``splitting'' (i.e., ``anticrossing'').

We found an analytical expression for the dispersion relations of the normal-mode oscillations for the case of a double circular chain of interacting particles, and we compared these relations with those for a double straight chain \cite{PRE_70}.
We showed that there are small differences between the dispersion relations of a circular and a straight both a single and a double chains.
These differences grow with increasing curvature of the parabolic confinement.

\section*{Acknowledgments}

We acknowledge fruitful discussions with F.~M.~Peeters, A.~Matulis and W.~P.~Ferreira. 
This work was supported by the ``Odysseus'' program of the Flemish Government and FWO-Vl.

\appendix
\renewcommand{\theequation}{\Alph{section}\arabic{equation}}

\begin{widetext}
\section{}

Here we present the analytical expressions for matrix elements of the dynamical matrix $\Lambda$ for a straight double chain (\ref{eqI}) discussed in Sec. III.
The corresponding independent matrix elements of the dynamical matrix $\Lambda$ are:

\begin{equation}
\label{eqI2}
\Lambda_{ll}^{xx} 
= \sum_{j=1}^\infty \frac{e^{-\kappa r}}{r^5} \left[\left(\left(\kappa r+1\right)^2 + 1\right) \left(j-\frac12\right)^2 a^2 - \left(\kappa r+1\right)d^2\right]\mid_{r=r_{-}} 
+ \frac{e^{-\kappa r}}{r^3}\left(\left(\kappa r+1\right)^2 + 1\right)\mid_{r=ja} \left(1-\cos\left(kja\right)\right)
\end{equation}

\begin{equation}
\label{eqI3}
\Lambda_{ll}^{yy}
= 1 + \sum_{j=1}^\infty \frac{e^{-\kappa r}}{r^5} \left[\left(\left(\kappa r+1\right)^2 + 1\right) d^2 - \left(\kappa r+1\right)\left(j-\frac12\right)^2 a^2\right]\mid_{r=r_{-}}
- \frac{e^{-\kappa r}}{r^3}\left(\kappa r+1\right)\mid_{r=ja} \left(1-\cos\left(kja\right)\right)
\end{equation}

\begin{equation}
\label{eqI4}
\Lambda_{ls}^{xx}=-\sum_{j=1}^\infty \frac{e^{-\kappa r}}{r^5} \left[\left(\left(\kappa r+1\right)^2 + 1\right) \left(j-\frac12\right)^2 a^2 - \left(\kappa r+1\right)d^2\right]|_{r=r_{-}} \cos\left(k\left(j-\frac12\right) a\right),\ \ \ l\neq s
\end{equation}

\begin{equation}
\label{eqI5}
\Lambda_{ls}^{yy}=-\sum_{j=1}^\infty \frac{e^{-\kappa r}}{r^5} \left[\left(\left(\kappa r+1\right)^2 + 1\right) d^2 - \left(\kappa r+1\right)\left(j-\frac12\right)^2 a^2\right]|_{r=r_{-}} \cos\left(k\left(j-\frac12\right) a\right),\ \ \ l\neq s
\end{equation}

\begin{equation}
\label{eqI6}
\Lambda_{ls}^{xy}=\Lambda_{ls}^{yx}
=i \sum_{j=1}^\infty \frac{e^{-\kappa r}}{r^5} \left[\left(\left(\kappa r+1\right)^2 + \left(\kappa r+1\right)+ 1\right)\right]|_{r=r_{-}} d\left(j-\frac12\right) a \sin\left(k\left(j-\frac12\right) a\right),\ \ \ l\neq s
\end{equation}

where $r_-=\sqrt{(j-\frac12)^2a^2+d^2}$ and $d$ is the equilibrium distance between the chains.
Other matrix elements can be calculated as hermitian conjugates using Eq. (\ref{eqHerm}).

\section{}
In this Appendix we present analytical expressions for matrix elements of the dynamical matrix $\Lambda$ (\ref{eqII}) discussed in Sec. IV for a circular double chain.
The independent matrix elements of the dynamical matrix $\Lambda$ for circular double chain are:

\begin{eqnarray}
\label{eqII1}
\Lambda_{ll}^{\phi\phi} 
& = & 
\sum_{j=1}^{
\left[
\frac{N}{4}\right]} \frac{e^{-\kappa r}}{r^5}\left[\left(\left(\kappa r+1\right)^2 + 1\right)\sin^2\left(\left(j-\frac12\right)\Delta\phi\right) 
\right. 
\\ \nonumber
& + & 
\left.
\left(\kappa r+1\right)\left(\left(\frac{r_{10}}{r_{20}}+\frac{r_{20}}{r_{10}}\right)\cos\left(\left(j-\frac12\right)\Delta\phi\right) - \left(1 + \cos^2\left(\left(j-\frac12\right)
\Delta\phi\right)\right)\right)
\right]
\mid_{r=\widetilde{r}_{0}}  
\\ \nonumber
& + & \sum_{j=1}^{\left[\frac{N}{4}\right]-1} r_{l0}^2\frac{e^{-\kappa r}}{r^5}
\left[\left(\left(\kappa r+1\right)^2 + 1\right)\sin^2\left(j\Delta\phi\right) + \left(\kappa r+1\right)\left(1-\cos\left(j\Delta\phi\right)\right)^2\right]|_{r=\overline{r}_{l0}} 
\\ \nonumber
& \times & 
\left(1-\cos\left(k r_0 j\Delta\phi\right)\right) 
\nonumber
\end{eqnarray}

\begin{eqnarray}
\label{eqII2}
\Lambda_{ll}^{rr}
& = & 
1+\sum_{j=1}^{
\left[\frac{N}{4}\right]} r_{10}r_{20}\frac{e^{-\kappa r}}{r^5}\left[
\left(\left(\kappa r+1\right)^2 + 1\right)\left(\left(\frac{r_{10}}{r_{20}}+\frac{r_{20}}{r_{10}}\right)\cos\left(\left(j-\frac12\right)\Delta\phi\right) 
\right.
\right.
\\ \nonumber
& - & 
\left.
\left.
\left(1 + \cos^2\left(\left(j-\frac12\right)\Delta\phi\right)\right)\right) 
+\left(\kappa r+1\right)\sin^2\left(\left(j-\frac12\right)\Delta\phi\right)\right]|_{r=\widetilde{r}_{0}}  
\\ \nonumber
& - & 
\sum_{j=1}^{\left[\frac{N}{4}\right]-1} r_{l0}^2\frac{e^{-\kappa r}}{r^5}
\left[\left(\left(\kappa r+1\right)^2 + 1\right)\left(1-\cos\left(j\Delta\phi\right)\right)^2 + \left(\kappa r+1\right)\sin^2\left(j\Delta\phi\right)\right]|_{r=\overline{r}_{l0}} 
\\ \nonumber
& \times & 
\left(1-\cos\left(k r_0 j\Delta\phi\right)\right) 
\nonumber
\end{eqnarray}

\begin{eqnarray}
\label{eqII3}
\Lambda_{ls}^{\phi\phi}
& = & 
-r_{10}^2r_{20}^2 \sum_{j=1}^{\left[\frac{N}{4}\right]} \frac{e^{-\kappa r}}{r^5}\left[\left(\left(\kappa r+1\right)^2 + 1\right)\sin^2\left(\left(j-\frac12\right)\Delta\phi\right) \right.
\\ \nonumber
& + & 
\left(
\kappa r+1\right)\left(\left(\frac{r_{10}}{r_{20}}+\frac{r_{20}}{r_{10}}\right)\cos\left(\left(j-\frac12\right)\Delta\phi\right) 
\right. 
\\ \nonumber
& - & 
\left. 
\left. 
\left(1 + \cos^2\left(\left(j-\frac12\right)\Delta\phi\right)\right)
\right)
\right]
\mid_{r=\widetilde{r}_{0}} \cos\left(k r_0 \left(j-\frac12\right) \Delta\phi\right),\ \ \ l\neq s
\end{eqnarray}

\begin{eqnarray}
\label{eqII4} 
\Lambda_{ls}^{rr}
& = & 
-r_{10}r_{20}\sum_{j=1}^{\left[\frac{N}{4}\right]} \frac{e^{-\kappa r}}{r^5}
\left[
\left(\left(\kappa r+1\right)^2 + 1\right)\left(\left(\frac{r_{10}}{r_{20}}+\frac{r_{20}}{r_{10}}\right)\cos\left(\left(j-\frac12\right)\Delta\phi\right) 
\right. 
\right. 
\\ \nonumber
& - & 
\left. 
\left. 
\left(1 + \cos^2\left(\left(j-\frac12\right)\Delta\phi\right)\right)
\right) 
\right.
\\ \nonumber
& + & 
\left. 
\left(\kappa r+1\right)\sin^2\left(\left(j-\frac12\right)\Delta\phi\right)\right]|_{r=\widetilde{r}_{0}} \cos\left(k r_0 \left(j-\frac12\right) \Delta\phi\right),\ \ \ l\neq s
\end{eqnarray}

\begin{equation}
\label{eqII5}
\Lambda_{ll}^{r\phi}=i r_{l0}^3 \sum_{j=1}^{\left[\frac{N}{4}\right]} \frac{e^{-\kappa r}}{r^5}
\left(1-\cos\left(j\Delta\phi\right)\right)\sin\left(j\Delta\phi\right)\left[\left(\kappa r+1\right)^2 + \left(\kappa r+1\right)+ 1\right]|_{r=\overline{r}_{l0}} \sin\left(k r_0 j\Delta\phi\right)
\end{equation}

\begin{eqnarray}
\label{eqII6}
\Lambda_{12}^{r\phi}
& = & 
i r_{20} \sum_{j=1}^{\left[\frac{N}{4}\right]} \sin\left(\left(j-\frac12\right)\Delta\phi\right) \frac{e^{-\kappa r}}{r^5}
\left[
\left(\kappa r+1\right)r_{20}\left(r_{20}-r_{10}\cos\left(\left(j-\frac12\right)\Delta\phi\right)\right)
\right.
\\ \nonumber
& + & 
\left. 
\left(\left(\kappa r+1\right)^2 + 1\right) r_{10}\left(r_{10}-r_{20}\cos\left(\left(j-\frac12\right)\Delta\phi\right)\right)  
\right]\mid_{r=\widetilde{r}_0} \sin\left(k r_0 \left(j-\frac12\right) \Delta\phi\right),\ \ \ l\neq s
\end{eqnarray}

\begin{eqnarray}
\label{eqII7}
\Lambda_{12}^{\phi r}
& = & 
-i r_{10} \sum_{j=1}^{\left[\frac{N}{4}\right]} \sin\left(\left(j-\frac12\right)\Delta\phi\right) \frac{e^{-\kappa r}}{r^5}
\left[
\left(\kappa r+1\right)r_{10}\left(r_{10}-r_{20}\cos\left(\left(j-\frac12\right)\Delta\phi\right)\right)
\right. 
\\ \nonumber
& + & 
\left. 
\left(\left(\kappa r+1\right)^2 + 1\right) r_{20}\left(r_{20}-r_{10}\cos\left(\left(j-\frac12\right)\Delta\phi\right)\right) 
\right]
\mid_{r=\widetilde{r}_0} \sin\left(k r_0 \left(j-\frac12\right) \Delta\phi\right),\ \ \ l\neq s
\end{eqnarray}

\noindent 
where $\overline{r}_{l0}=\sqrt{2} r_{l0}\sqrt{1-\cos\left(j\Delta\phi\right)}$ and $\widetilde{r}_0=\sqrt{r_{10}^2+r_{20}^2- 2r_{10}r_{20}\cos\left(\left(j-\frac12\right)\Delta\phi\right)}$.
Other elements of the dynamical matrix can be found in a similar way using Eq.~(\ref{eqHerm}).

The asymptotic relations in the limit $r_{l0}\rightarrow\infty,\ (\Delta\phi\rightarrow0)$ and $\frac{d}{\sigma}\rightarrow0$,

\begin{equation}\label{eqII8}
\sqrt{r_{10}^2+r_{20}^2- 2r_{10}r_{20}\cos\left(\left(j-\frac12\right)\Delta\phi\right)} \rightarrow \sqrt{\left(j-\frac12\right)^2a^2+d^2},
\end{equation}
\begin{eqnarray}
\sqrt{2} r_{l0}\sqrt{1-\cos\left(j\Delta\phi\right)} \rightarrow ja,\ \nonumber
|r_{l0}-r_{s0}\cos\left(\left(j-\frac12\right)\Delta\phi\right)| \rightarrow d,\ l\neq s,\
\end{eqnarray}
$$
r_{10}r_{20}\left(\left(\frac{r_{10}}{r_{20}}+\frac{r_{20}}{r_{10}}\right)\cos\left(\left(j-\frac12\right)\Delta\phi\right) - \left(1 + \cos^2\left(\left(j-\frac12\right)\Delta\phi\right)\right)\right) \rightarrow d^2,
$$
$$
r_{10}r_{20}\sin^2\left(\left(j-\frac12\right)\Delta\phi\right) \rightarrow \left(j-\frac12\right)^2 a^2.
$$
\end{widetext}

\end{document}